\documentclass[a4paper]{article}
\begin{document}
\title{Generalized Quantum Search Hamiltonian}
\author{Joonwoo Bae{\footnote{Email address: jwbae@newton.hanyang.ac.kr}}, Younghun Kwon{\footnote{ Email address: yhkwon@newton.hanyang.ac.kr}} \\Department of Physics, Hanyang University, \\Ansan, Kyunggi-Do, 425-791, South Korea}
\date{\today}
\maketitle 
\begin{abstract}
There are hamiltonians that solve a search problem of finding one of $N$ items in $O(\sqrt{N})$ steps. They are hamiltonians to describe an oscillation between two states. In this paper we propose a generalized search hamiltonian, $H_{g}$. Then the known search hamiltonians become special cases of $H_{g}$. From the generalized search hamiltonian, we present remarkable results that searching with $100 \%$ is subject only to the phase factor in $H_{g}$ and independent to the number of states or initialization.
\end{abstract}
 
 Grover proposed the quantum search algorithm that finds one of unsorted $N$ items in $O(\sqrt{N})$ steps.\cite{grover} Since it is known that classical search algorithms would take $O(N)$ steps to solve a search problem, the quadratic speedup in Grover algorithm is remarkable. The reason of the speedup is effects of quantum mechanics. To apply quantum mechanics to a search problem, it needs to map each item to a state, respectively, in $N$ dimensional Hilbert space. Then a search problem can be transformed to finding one of $N$ states. We can use quantum mechanical characteristics, for instance, superposition and parallelism. Grover algorithm is in fact to find the state which corresponds to a target item. An iteration of Grover algorithm is composed of two operations: the first is to flip the target state about $0$ and the next to invert all states about the average state.\cite{jozsa} An oracle, the heart of quadratic speedup, is applied in the first operation. Grover iterations amplify the amplitude of the target state up to nearly one in $O(\sqrt{N})$ times, when an initial state is a uniform superposition of $N$ states. Zalka showed Grover algorithm is optimal when unstructured items are considered.\cite{zalka} \\
 On the other hand, there are a search algorithm based on hamiltonian evoution by Schrodinger equation. While Grover algorithm operates a state  in discrete time, a search hamiltonian does a state in continuous time. They are hamiltonians to describe an oscillation between two states. Farhi et al. suggested harmonic-oscillation hamiltonian, exactly $H_{fa} = E(|w \rangle \langle w|+|s \rangle \langle s| )$, where $|w \rangle$ is a target state and $|s \rangle$ is an initial state.\cite{farhi} Fenner provided another hamiltonian, $H_{fe}=2iEx(|w \rangle \langle \psi|-|\psi \rangle \langle w|)$, where $x$ is the overlap between an initial state and a target state, i.e., $ \langle w | s \rangle = x (>0) $.\cite{fenner} A state under the hamiltonians evolves from an initial state to a target state in the half period of oscillation, $O(\sqrt{N})$. This oscillation is a process to amplify the amplitude of a target state. It is noted that the initial state should be prepared as a superposition of all states, since one of initially superposed states becomes a target state by the hamiltonian evolution.\\
 We expect that there exists a more general search hamiltonian including the known hamiltonians. We then write down a most possible combination of a target state and an initial state as follows:

\begin{displaymath}
H=E(a |w \rangle \langle w| + b |w \rangle \langle s| + c |s \rangle \langle w| + d|s \rangle \langle s|)
\end{displaymath}

The initial state can be written as $|s \rangle = x |w \rangle + \sqrt{1-x^2}|r \rangle $, where $|r \rangle$ is the state orthogonal to the target state(so $\langle w|r \rangle=0$) and $x$ is the overlap between the initial state and the target state, i.e., $x=\langle w|s \rangle $. The hamiltonian and the initial state can be written in matrix representation: 

\begin{displaymath}
\mathbf{H} = E \left(  \begin{array}{ccc} a+(b+c)x+dx^2 & (b+dx)\sqrt{1-x^2} \\ (c+dx)\sqrt{1-x^2} & d(1-x^2) \end{array} \right), 
\end{displaymath}
\begin{displaymath}
\mathbf{|s \rangle } = 
\left( \begin{array}{ccc} x \\ \sqrt{1-x^2} \end{array} \right)
\end{displaymath}

A state $| \psi(t) \rangle$ is governed by time-dependent Schrodinger equation:

\begin{displaymath}
i\frac{d}{dt}|\psi(t) \rangle =H|\psi(t) \rangle
\end{displaymath}

where the initial condition is $|\psi(t=0) \rangle=|s \rangle$(we let $\hbar=1$ for convenience). After t times the probability to find the target state is :

\begin{eqnarray}
\lefteqn{P   =  {|\langle w|{e}^{-iHt}|s \rangle|}^{2} } \nonumber \\ 
    & =&  x^{2} cos^{2} EDt +  C C^{*}sin^{2}EDt
\end{eqnarray}

where, $q=\frac{1}{2}((a+d)+(b+c)x)$, \\ $D=\sqrt{q^2 - (ad-bc)(1-x^2)}$, and $C=$
\begin{displaymath}
[\frac{{(d(1-x^2)-q)}^{2}+x(c+dx)(d(1-x^2)-q) - D^2 }{(c+dx)D}] 
\end{displaymath}

 The postulate of quantum mechanics, that any observable should be hermitian, requires $H=H^{\dag}$. Then we obtain : \\ [2pt]

1. $a$ and $d$ are real. \\ [2pt]

2. $b=c^{*}=r e^{i \phi}$, where r and $\phi$ are real. \\ [2pt]

 Suppose the initial state is a uniform superposition of all states and we read out the final state at time $T$. In equation $(1)$, the contribution of $x^{2} cos^{2}EDT$ to the probability is so small that it can be negligible. Moreover, $x^{2} cos^{2}EDT$ becomes zero at time $T$, since the read-out time $T$ comes from $EDT=\frac{\pi}{2}$. We find out $T=O(\sqrt{N})$, where the quadratic speedup is shown. Thus it is essential to make the value of $C^{*}C$ as large as possible, to find the target state with high probability at time $T$.

\begin{eqnarray}
\lefteqn{P   =  C C^{*} } \nonumber \\ 
   & = & \frac{u_{0}+u_{1} x+u_{2} x^{2}+u_{3} x^{3}+u_{4} x^{4}+u_{5} x^{5}+u_{6} x^{6}}{l_{0}+l_{1}x +l_{2} x^{2}+l_{3} x^{3} +l_{4} x^{4}} \nonumber \\ 
 & &  \\
u_{0} &=& r^{4}  \nonumber\\ 
u_{1} &=& (a+3d)r^{3} cos \phi  \nonumber\\ 
u_{2} &=&  \frac{1}{4}r^{2}(a^{2} +6ad +9d^{2}-4r^{2} \nonumber\\ && +4(ad+d^{2}+r^{2}) cos 2\phi) \nonumber\\ 
u_{3} &=& \frac{1}{2}dr cos \phi (a^{2} +4ad+3d^{2} -4r^{2}+4r^{2}cos2\phi) \nonumber\\ 
u_{4} &=& \frac{1}{4} (a^{2} d^{2} +2ad^{3} +d^{4} -4d^{2} r^{2} +2r^{4} \nonumber\\ && +(4d^{2} r^{2}- 2r^{4})cos2\phi) \nonumber\\ 
u_{5} &=& 2d r^{3} cos\phi sin^{2} \phi  \nonumber\\ 
u_{6} &=& d^{2} r^{2} sin^{2} \phi  \nonumber\\ 
l_{0} &=& \frac{1}{4}(a^{2} r^{2} -2adr^{2} + d^{2} r^{2} + 4r^{4} ) \nonumber\\ 
l_{1} &=& \frac{1}{2}r (a^{2} d -2ad^{2} +d^{3}+2ar^{2} + 6dr^{2} )cos\phi  \nonumber\\ 
l_{2} &=& \frac{1}{4} (a^{2} d^{2} +d^{4} +8d^{2}r^{2} -2r^{4} -2a(d^{3}-4dr^{2}) \nonumber\\ && + 2r^{2}(2ad +2d^{2} +r^{2})cos 2\phi)  \nonumber\\ 
l_{3} &=& dr cos \phi (3ad +d^{2} - r^{2} +r^{2} cos2\phi )  \nonumber\\ 
l_{4} &=& \frac{1}{2}d^{2}(2ad - r^{2} + r^{2} cos2\phi)  \nonumber\\ 
\end{eqnarray}

 In (2), to consider the zeroth order terms of $x$, $u_{0}$ and $l_{0}$, is crucial. Thus we obtain a condition $a=d$ by setting $u_{0}=l_{0}$, which automatically gives $u_{1}=l_{1}$. Under the condition $a=d$, the probability becomes:

\begin{eqnarray}
P_{g}=1-O(x^{2})
\end{eqnarray}

The corresponding hamiltonian is:

\begin{eqnarray}
\lefteqn{ H_{g}=E_{1}(|w \rangle \langle w| + |s \rangle \langle s|) } \nonumber\\ && + E_{2}(e^{i\phi}|w \rangle \langle s| + e^{-i \phi} |s \rangle \langle w|) 
\end{eqnarray}

where $E_{1}=Ea$ and $E_{2}=Er$. The probability to get the target state under this hamiltonian is not exactly but nearly one. Therefore it is allowed to use hamiltonian $H_{g}$ when searching with $100 \%$ is not required.  \\ 
 We here pursue condition for higher probability, which is $l_{2}=u_{2}$. It makes us choose a specific phase, $\phi = n \pi $. This automatically gives useful relations, $l_{3}=u_{3}, l_{4}=u_{4}$, and $u_{5}=u_{6}=0$. Then the probability under these relations is:

\begin{displaymath}
P_{p}=\frac{u_{0}+u_{1} x+u_{2} x^{2}+u_{3} x^{3}+u_{4} x^{4}}{u_{0}+u_{1}x +u_{2} x^{2}+u_{3} x^{3} +u_{4} x^{4}}=1
\end{displaymath}

The hamiltonian is :

\begin{eqnarray}
H_{p}=E_{1} (|w \rangle \langle w| + |s \rangle \langle s|) \pm E_{2} (|w \rangle \langle s| + |s \rangle \langle w|) 
\end{eqnarray}

 That is, hamiltonian $H_{p}$ can search a target state with probability one. \\

 We now have the generalized search hamiltonian, $H_{g}$, and the perfect search hamiltonian $H_{p}$. $H_{p}$ is in particular produced from $H_{g}$, by fixing the phase $\phi=n \pi$. Here, we can show that known search hamiltonians are special forms of $H_{g}$. For $E_{2} =0$, it becomes Farhi's hamiltonian, and Fenner's hamiltonian for $E_{1}=0$, $\phi = \frac{\pi}{2}$, and $r=2x$. Moreover, we provide a new quantum search hamiltonian :

\begin{eqnarray}
H_{new}=  E_{2}(e^{i\phi}|w \rangle \langle \psi| + e^{-i \phi} |\psi \rangle \langle w| )
\end{eqnarray}

 When Grover algorithm is applied to a search problem, the probability one appears if and only if two assumptions are satisfied : 1) the target items are $\frac{N}{4}$ among $N$ items, 2) the initial state is a uniform superposition of all states in a system. Here we present the remarkable result that hamiltonian $H_{p}$ does not depend on the number of states and initialization. There is no condition on $H_{p}$ about the number of states or initialization. This implies that $H_{p}$ finds target states with probability one regardless of the number of target states, though the initial state is prepared as an arbitrary superposition of all states. When a generalized search hamiltonian is used in a search problem, probability one at read-out procedure depends only on the phase condition.

\section*{Acknowledgement}
J. Bae is supported in part by the Hanyang University Fellowship and Y. Kwon is supported in part by the Fund of Hanyang University.

\end{document}